\documentstyle[12pt,epsfig]{article}
\begin{document}
\normalsize \rm

\begin{center}
{\bf \LARGE Self-compensating incorporation of Mn in
Ga$_{1-x}$Mn$_{x}$As}
\\[16pt]

{J. Ma\v{s}ek and F. M\'{a}ca}
\\
Institute of Physics, Academy of Sciences of the CR\\ CZ-182 21
Praha 8, Czech Republic\\
\end{center}
\vspace{22mm}

\begin{abstract}
We consider hypothetical Ga$_{7}$MnAs$_{8}$, Ga$_{16}$MnAs$_{16}$,
and Ga$_{14}$Mn$_{3}$As$_{16}$ crystals with Mn in a
substitutional, interstitial, and both positions. Spin-polarized
FPLAPW calculations were used to obtain their electronic
structure. We show that the interstitial Mn acts as a double donor
and compensates the holes created by two Mn atoms in
substitutional positions. This explains why the number of holes in
Ga$_{1-x}$Mn$_{x}$As is much smaller than x. The presence of
interstitial atoms may also be the reason for the lattice
expansion with increasing content of Mn. The differences in
electronic behavior of substitutional and interstitial Mn are
discussed.
\\


\end{abstract}

\section{Introduction}
The III-V diluted magnetic semiconductors containing Mn attracted
much attention in last years \cite{R1,R2} because their
ferromagnetic nature is promising for applications in
semiconductor structures. The ferromagnetic behavior of III-V
diluted magnetic semiconductors (DMS) is connected with their
p-type nature \cite{R1}. Mn substituted for a trivalent cation
acts as an acceptor and creates a hole in the valence band. If the
content of Mn is of order of one percent, the Fermi energy is
fixed in the valence band and all other defects and impurities are
less important. The main contribution to the exchange interaction
between Mn local moments is then mediated by the holes with wave
vectors close to the center of the Brillouin zone. The typical
period of the RKKY interaction exceeds the average Mn - Mn
distance and the coupling is ferromagnetic. The correlation
between magnetic and transport measurements \cite{R2} supports
this picture.

Recently, also the electronic structure of Mn-doped III-V
compounds has been investigated. Both supercell band structure
calculations \cite{R3,R4} and KKR-CPA studies \cite{R5} confirmed
the p-type character of these materials as well as the presence of
localized magnetic moments at Mn sites, and their ferromagnetic
coupling.

However, there are still some unclear points. One of them is a
remarkable difference between the number of Mn acceptors and the
number of free holes obtained from the transport measurements. The
latter quantity is usually much smaller than the former. To
explain this almost complete compensation, it is assumed that most
of the holes do not participate in the conduction because they are
either tightly bound to the acceptors \cite{R6}, localized due to
the disorder, or compensated by As antisite defects \cite{R7}.

We propose an alternative explanation of the self-compensation
property of the Mn impurities.
We assume that some of Mn atoms do not substitute into the cation sublattice,
but occupy interstitial position in the zinc blende structure.
One can expect that the interstitial Mn acts as a double donor,
because there is no space for its two 4s electrons in the bonding.
If it is so, than one interstitial Mn should compensate the holes
created by two substitutional atoms.

To check this idea, we constructed a series of hypothetical
crystals whose large unit cells consist of a few conventional
cubic cells of GaAs and contain Mn in either substitutional or
interstitial positions. We calculated the electronic structure for
these superstructures and found the positions of the Fermi level
with respects to the valence and conduction bands. The Fermi level
lying in the valence band indicates that the impurity behaves as
an acceptor and the number of empty states per unit cell defines
its degree of ionization. Similarly, the donor case can be
recognized according to partly occupied conduction band. In
addition, we investigated also a Ga$_{14}$Mn$_{3}$As$_{16}$
crystal with one interstitial and two substitutional Mn atoms to
approach the real charge distribution in a compensated case.
Although the content of Mn is strongly overestimated in this case,
we use it to show the differences in the local electronic
structure at substitutional and interstitial atoms.

\section{Details of calculation}
The self-consistent, spin-polarized electronic structure of all
considered systems was calculated by means of the full-potential
linearized augmented plane wave (FPLAPW) method \cite{R8}. Instead
of the standard form of the density functional we used the GGA
version, giving a wider band gap and a better description of the
conduction band. All calculations were done for a ferromagnetic
phase.

We started with the band structure of the host GaAs crystal.
The calculated band gap is 0.57 eV.
Although this value is better than the gap obtained by using
the local-density approximation (0.43 eV),
it is still underestimated with respect to observed value 1.56 eV.
A comparable relation between calculated and observed values of the gap
can be expected also for the systems containing Mn.

In our calculations we use a large unit cell (LUC) consisting of
four cubic cells of the zinc-blende structure and containing 16
molecular units of the host GaAs. It is constructed by doubling
the c-axis and by assuming a $\sqrt(2) \times \sqrt(2)$
superstructure in the basal plane. This LUC has a tetragonal
symmetry and its lattice parameters are $A = B = a \times
\sqrt(2), C = 2 \times a$, with $a = 0.566 + 0.032 \times x$ (nm)
\cite{R9} being the lattice constant of the Ga$_{1-x}$Mn$_{x}$As.
We do not take into account any local relaxation around Mn
impurities, i.e. the mixed systems are considered with a perfect
tetrahedral bonding. In this way, both interstitial and
substitutional Mn have the same geometry of the nearest neighbor
sphere. We use the convention with the basal plane coinciding with
the cation layer.

\section{Results}
\subsection{Mn in a substitutional position}

We replace two Ga atoms of the GaAs LUC by Mn. Two impurities are
considered because we presume that this is the number of
substitutional Mn atoms, which can be compensated by a single
interstitial Mn. The two Mn atoms are placed at the most distant
positions in the LUC. The choice of relative coordinates (1/2,
1/2, 1/4) and (1/2, 1/2, 3/4) does not change the tetrahedral
symmetry of the LUC. Moreover, this unit cell of
Ga$_{14}$Mn$_{2}$As$_{16}$ can be decomposed into two equal cells
Ga$_{7}$MnAs$_{8}$, which were used in the actual calculations for
this model.

\begin{figure}[tbp]
\begin{center}
\epsfig{file=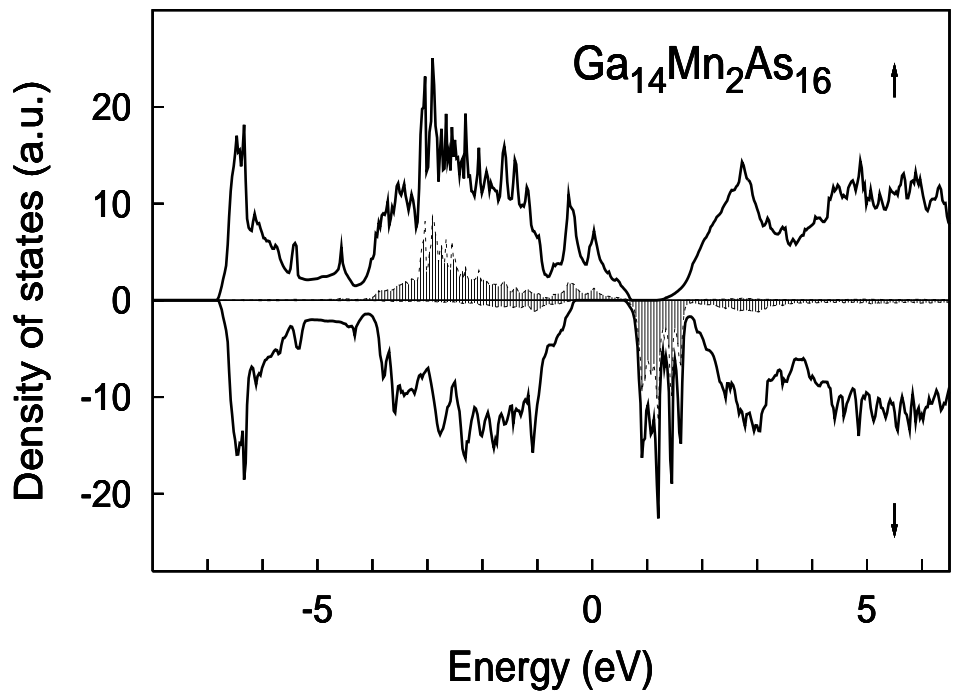, height=13.5cm, width=9cm, angle=270}
\end{center}
\caption{Spin-polarized density of states for
Ga$_{14}$Mn$_{2}$As$_{16}$ crystal with Mn atoms in substitutional
positions. The shaded areas show the contribution of Mn 3d states.
Vertical line indicates the position of the Fermi level.}
\end{figure}

The resulting spin-polarized density of states (DOS) is shown in Fig. 1.
We see that the Ga$_{14}$Mn$_{2}$As$_{16}$ crystal is a semimetal with
the Fermi energy close to the center of a wide band gap (0.82 eV)
for the minority-spin electrons.
For the majority-spin electrons, however, the bands derived
from the host valence and conduction bands are separated by a narrow gap (0.11 eV).
The Fermi energy crosses the valence band 0.75 eV bellow the top,
leaving just one state per unit cell empty.
This agrees with the previous findings [6-8] that substitutional Mn
in GaAs acts as an acceptor.
The spin of the hole has a sign opposite to the local moment
arising from the saturated spin polarization of the Mn 3d states.
As a result, the total spin of the formula unit
(which is assigned to two Mn) is just 4.
The integer value of the spin is related to the absence
of a free Fermi surface for the minority-spin.

The main spectral feature due to the presence of Mn is the
appearance of the d-states in the valence and conduction band. The
occupied majority-spin d-states contribute to the valence band
spectrum in a much wider range of binding energies (1 - 4 eV) than
in II-VI's. There is also a remarkable mixing of the minority-spin
d-states with the lowest conduction bands of the host crystal. The
bottom of the conduction band is formed of d-states instead of
cation s-states, as it is usual in the zinc-blende materials. The
average exchange splitting of the d-states is reduced from approx.
6 eV typical for Mn to approx.  4 eV. This underestimation of the
exchange splitting is the result of the density-functional
approach and, in fact, we should expect the empty Mn d-states
located at a higher energy.

Another important change of the electronic spectrum
of Ga$_{1-x}$Mn$_{x}$As concerns the cationic s-states.
While the Ga s-states are known to contribute mostly to
the bottom of the conduction band and to the lower valence band
(4 - 7 eV bellow the Fermi level), the Mn s-states appear,
both in valence and conduction bands, at much higher energies.
The shift is approx. 4 eV.
This large difference in the atomic level positions of the host
and substituted atoms represents, besides the presence of the d-states,
another important channel for the scattering of band carriers in the mixed crystal.
The strong alloy scattering can be also expected to affect
the near-edge optical transitions.

\subsection{Mn in the interstitial position}
The most probable position for a metallic interstitial in the
zinc-blende structure is a tetrahedral hollow site surrounded by
four anions. If we place the Mn atom to the (0, 0, 1/2) position
in the LUC, the resulting structure of Ga$_{16}$MnAs$_{16}$ has
again the complete tetragonal symmetry. The interstitial Mn has,
in addition to four As nearest neighbors at bonding distances $d1
= a \times \sqrt(3) / 4$, six close Ga neighbors at a distance $d2
= 1.154 \times d1$.

We find that Ga$_{16}$MnAs$_{16}$ is metallic, with the Fermi energy
in the conduction band for both spin directions.
The identity of the host valence and conduction bands can be still recognized.
As in the substitutional case, the gap for the majority-spin electrons
is much narrower (0.14 eV) than the band gap for the minority-spin electrons (0.37 eV).
There are, altogether, two electrons in the conduction band,
which confirms the idea that the interstitial Mn in GaAs acts as a double donor.

The total spin of the unit cell is 1.56.
Its main part  (1.35) is localized in the muffin-tin sphere at the Mn site.
This local moment is so small because also minority-spin d-states,
participating in the lowest conduction band, are partly occupied.
The remaining magnetization is distributed over the neighboring As atoms
and in the interstitial space around Mn.

The reduced magnetic moment of Mn is closely related
with a corresponding reduction of the average exchange splitting
of the Mn d-states (approx. 2.5 eV).
This has a pronounced effect on the distribution of the
d-states in the valence band.
In contrast to the substitutional case, the d-states shrink
to a narrow peak in the uppermost part of the valence band.

\subsection{Compensated case}
Finally, we consider a Ga$_{14}$Mn$_{3}$As$_{16}$ crystal
containing Mn in both substitutional and interstitial positions.
We assume that the two substitutional impurities and one
interstitial Mn have the same, high symmetry positions as in the
structures discussed in Sect. 3.1. and 3.2.. In this way, the
smallest Mn - Mn distance equals to the lattice constant a, i.e.
it corresponds to the next nearest neighbors in the cationic
sublattice.
Although the concentration of Mn in this model is largely overestimated (19%)
the closest Mn - Mn pairs are avoided and the system can still be
considered to be dilute.

The density of states presented in Fig. 2 indicates the
compensation. The position of the Fermi energy is fixed in the
minimum of the total DOS. Analyzing the band structure, we found
an overlap of the majority-spin valence band ($E_{v}(\uparrow) =
0.37 $eV) with the minority-spin conduction band
($E_{c}(\downarrow) = -0.35 $eV). This means that there are some
holes in majority-spin valence band and the same amount of
minority-spin conduction electrons. Their concentration is 0.562
per unit cell, i.e. 0.187 per Mn atom, much less than in the case
of either substitutional or interstitial doping.

\begin{figure}[tbp]
\begin{center}
\epsfig{file=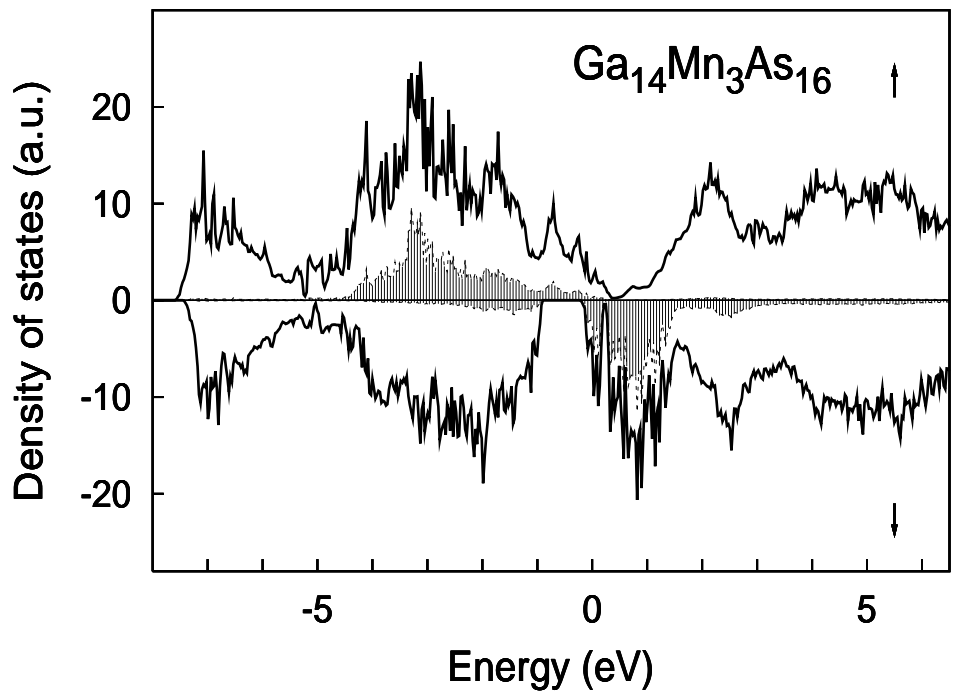, height=13.5cm, width=9cm, angle=270}
\end{center}
\caption{Same as Fig. 1 but for Ga$_{14}$Mn$_{3}$As$_{16}$ crystal
with one interstitial and two substitutional Mn atoms.}
\end{figure}

It is important to notice that all these numbers depend very
sensitively on the relative position of $E_{v}(\uparrow)$ for the
majority- and $E_{c}(\downarrow)$ for the minority-spin electrons.
The number of free carriers decreases to zero with decreasing
$E_{v}(\uparrow) - E_{c}(\downarrow)$. This overlap,
characterizing the semimetallic state of
Ga$_{14}$Mn$_{3}$As$_{16}$, is due to the strong magnetic
polarization of the bands, which is clearly overestimated in our
example with so high content of Mn. In real mixed crystals with
low concentration of Mn, the overlap of the valence and conduction
band will be much smaller and the compensation better. A linear
interpolation indicates that the overlap disappears for
Ga$_{1-x}$Mn$_{x}$As with $x < 0.04$ and with this proportion
(2:1) of substitutional and interstitial Mn. The material then
behaves as a compensated semiconductor.

Although the geometry of both substitutional and interstitial Mn
is the same within the nearest-neighbor sphere,
the bonding and the local electronic structure are quite different.
The main difference concerns the local density of d-states.
We find that the occupied d-states at the interstitial Mn
approach the top of the valence band.
Even though the effect is not so strong as in Sect. 3.2.,
the change of  the d-states DOS should be visible e.g. in the X-ray emission.

\section{Conclusios}
Using ab-initio FPLAPW calculations, we showed
that the interstitial Mn in GaAs acts as a double donor.
Such impurity compensates the effect of two Mn atoms substituted
into the cation sublattice, which are known to be single acceptors.
The number of holes in the valence band becomes much smaller
than the total number of magnetic impurities,
if only a minor portion of them occupies interstitial instead of
substitutional positions.
This can explain the large difference between the level of Mn doping and
the resulting hole concentration observed in experiment [7].

The presence of interstitial Mn in Ga$_{1-x}$Mn$_{x}$As might be verified
by using the X-ray emission spectra from Mn.
The L spectra, corresponding to the local density of Mn d-states,
should show a remarkable difference for the two bonding geometries in question.
The main reason for this difference is the reduced exchange splitting
of the d-states at the interstitial Mn, which brings the occupied
d-states into the uppermost part of the valence band.
\\

\noindent
{\bf Acknowledgement:}
Financial support for this work was provided by the COST project P3.80.
\\

\end{document}